\documentclass{birkjour}
\newtheorem{theorem}{Theorem}[section]

\newtheorem{lemma}[theorem]{Lemma}

\newtheorem{proposition}{Proposition}

\theoremstyle{definition}
\newtheorem{definition}[theorem]{Definition}
\newtheorem{remark}{Remark}

\newtheorem{example}{Example}

\newcommand{\al}{\mathfrak{g}}

\newcommand{\R}{\mathbb{R}}      
\newcommand{\ra}{\rightarrow}

\newcommand{\tpair}{\widetilde{T\langle\cdot,\cdot\rangle}}
\makeatletter
\def\lcf{\lbrack\! \lbrack}
\def\rcf{\rbrack\! \rbrack}

\def\lcf{\lbrack\! \lbrack}
\def\rcf{\rbrack\! \rbrack}

\begin{document}

%
%
%
%
%
%
%
%
%

\title[Second-order mechanical systems on Lie algebroids]{Lagrangian Lie subalgebroids generating dynamics for second-order mechanical systems on Lie algebroids}

\author{L\'{i}gia Abrunheiro}

\address{Center for Research and Development in Mathematics and Applications\\ 
ISCA, Universidade de Aveiro,\\
3810-500 Aveiro, Portugal}
\email{abrunheiroligia@ua.pt}

\author[Leonardo Colombo]{Leonardo Colombo$^{*}$}
\address{Instituto de Ciencias Matem\'aticas (CSIC-UAM-UC3M-UCM)\\
C/Nicol\'as Cabrera 13-15, 28049.\\
Madrid, Spain}

\email{leo.colombo@icmat.es}

\thanks{$*$ corresponding author. Send correspondence about the article to leo.colombo@icmat.es. The work of Leonardo Colombo was partially supported by Ministerio de Econom\'ia, Industria y
Competitividad (MINEICO, Spain) under grant MTM2016-76702-P and ''Severo Ochoa Programme for Centres of Excellence''
in R$\&$D (SEV-2015-0554). The work of L\'{i}gia Abrunheiro was supported by Portuguese funds through the \emph{Center for Research and Development in Mathematics and Applications} (CIDMA)  and the \emph{Portuguese Foundation for Science and Technology} (``FCT--Funda\c{c}\~ao para a Ci\^encia e a Tecnologia''), within project UID/MAT/04106/2013. Part of the results of this work corresponds with the Ph.D thesis of the corresponding author \cite{thesis}. }




\begin{abstract}
The study of mechanical systems on Lie algebroids permits an understanding of the dynamics described by a Lagrangian or Hamiltonian function for a wide range of mechanical systems in a unified framework. Systems defined in tangent bundles, Lie algebras, principal bundles, reduced systems and constrained are included in such description. 

In this paper, we investigate how to derive the dynamics associated with a Lagrangian system defined on the set of admissible elements of a given Lie algebroid using Tulczyjew's triple on Lie algebroids and constructing a Lagrangian Lie subalgebroid of a symplectic Lie algebroid, by building on the geometric formalism for mechanics on Lie algebroids developed by M. de Le\'on, J.C. Marrero and E. Mart\'inez on ``Lagrangian submanifolds and dynamics on Lie algebroids''. 
\end{abstract}

\maketitle
\section{Introduction}
Lie algebroids have deserved a lot of interest in the last decades in the field of Geometric Mechanics since these spaces generalize the traditional framework of tangent bundles to more general situations (Lie algebras, principal bundles, semi-direct products), including, for instance, systems with symmetries and constrained \cite{GU}, \cite{IMMS}, \cite{MdLMdDJCM}, \cite{JCsolo}. If a Lagrangian specifying the dynamics of a mechanical systems  is given, and it is invariant under a Lie group of symmetries, the description of the system relies on the geometry of a quotient space under the action of the Lie group and the dynamics is governed by Lagrange-Poincar\'e equations. In that sense, Weinstein \cite{We} showed that the common geometrical structure of the Lagrange-Poincar\'e equations in mechanics is the same as the one for Euler-Lagrange equations, a Lie algreborid structure.  

In the recent years, the study of higher-order mechanics, and in particular second-order mechanics, have been growing by their relevant applications to interpolation problems, optimal control, trajectory planning of vehicles, computational anatomy among others \cite{ACCCMS}, \cite{Pepin2007}, \cite{Colombo}, \cite{ldm}, \cite{CoMdD}, \cite{MR2580471}, \cite{DavidM}. The geometrical description of higher-order mechanical systems on Lie algebroids have been recently studied by the interest in the applications mentioned before and its subjacent geometry \cite{Bruce1}, \cite{Bruce3}, \cite{pichon1}, \cite{pichon3}, \cite{Eduardoho}. 

Part of the question how to deal with second order systems on Lie algebroids is what they really are. In the literature there are essentially two concepts of higher order mechanical systems on Lie algebroids. The first one, due to Mart\'inez and collaborators, uses the language of Lie algebroid prolongations both for first and higher order systems \cite{Eduardo}, \cite{Eduardoho}; the second, due to Grabowski and collaborators
\cite{Bruce1}, \cite{Bruce3}, \cite{GU} uses the concept of Tulczyjew triple and geometry of graded bundles  (see \cite{MJprolongations} for a recent work attending this question). In this work we study the unification of both approaches of higher-order mechanics on Lie algebroids for second-order Lagrangian systems into a symplectic framework by understanding the derivation of the dynamics using a Lagrangian Lie subalgebroid of a symplectic Lie algebroid based, and building, on the work done by de Le\'on, Marrero and Mart\'inez in \cite{LMM}. In such a work the authors propose an interpretation of the equations of motions based in the construction of a Tulczyjew triple in prolongations of Lie algebroids which consists of three prolongations of Lie algebroids retaining the dynamics into a symplectic framework that does not depends on the choice of the Lagrangian and its regularity, and moreover, it connect all the symplectic structures of the symplectic Lie algebroids that appear in the triple by symplectomorphisms. 

The ideas behind this triple have been extended to other general structures different than Lie algebroids as Dirac structures \cite{GG} and to more general classes of systems such as field theories \cite{CGM}, \cite{G}, Lie groups \cite{Ogul1}, \cite{Ogul2} and reduced systems \cite{EduardoGT}, \cite{ZG}.

The main source of inspiration for the idea behind the construction in this paper is the construction for the tangent bundle version developed by de Le\'on and Lacomba \cite{LeonLacomba} using Lagrangian submanifolds to obtain the dynamics associated with higher-orger Lagrangian systems. This work is also inspired by recent developments on mechanics on a prolongation Lie algebroids for the construction of geometric numerical methods \cite{DavidRodrigo}. 

In Section $2$ we recall the geometric structures we will use in the work, more precisely, prolongations of Lie algebroids, symplectic Lie algebroids, Lie subalgebroids, and  the set of admissible elements. Next, in Section $3$ we introduce a Lagrangian Lie subalgebroid and the extension of the Tulczyjew triple to prolongation of Lie algebroids. Based on those objects we derive the corresponding dynamics for second-order Lagrangian systems. Some examples are explored in Section $4$ to show how to interpret the dynamics of second-order Lagrangian systems in the proposed framework.

\section{Preliminaries}\label{s2}
Let $A$ be a Lie algebroid of rank $n$ over a smooth  manifold $M$ of dimension $m$. That is, $A$ is  a real vector bundle $\tau_A:A\rightarrow M$ together with a Lie bracket $\lcf\cdot,\cdot
\rcf$ defined on the $C^{\infty}(M)$-module $\Gamma(\tau_A)$ of sections of $A$, and a fiber map $\rho:A\ra TM$ called \textit{anchor map}.

Along the paper, we denote by $(x^{i})$, $i=1,\ldots,n$, local coordinates on an open subset $U$ of $M$. Given a local basis $\{e_{\alpha}\}$ of sections of $A$, $\alpha=1,\ldots,m$; $\rho_{\alpha}^{i}$ denotes the components of the anchor map $\rho$ with respect to this basis. The basis $\{e_{\alpha}\}$  together with the local coordinates $(x^{i})$ induces local coordinates $(x^i, y^{\alpha})$ on $A$ where each $a\in A$ is given in such a basis as $e=y^1e_1(\tau_A(a)) +\ldots+y^me_m(\tau_A(a))$.

\subsection{Prolongation of a Lie algebroid}

The \textit{prolongation of a Lie algebroid over a smooth map} $f:M'\rightarrow M$ is defined by
$$
\mathcal{T}^{f}A:=\bigcup_{m\in M'}A_{\rho}\times_{Tf}T_{m}M'=\bigcup_{m\in M'}\{(b,v')\in A\times T_{m}M'\mid\rho(b)=(Tf)(v')\},
$$
where  $Tf:TM'\rightarrow TM$ denotes the tangent map to  $f.$ In particular, the prolongation of a Lie algebroid $A$ over its canonical projection $\tau_A:A\rightarrow M$, is given by
$$
E:=\mathcal{T}^{\tau_A}A=\bigcup_{a\in A}A_{\rho}\times_{T\tau_A} T_{a}A=\bigcup_{a\in A}\{(b,v_a)\in A\times
T_{a}A\mid\rho(b)=(T_{a}\tau_A)(v_a)\}.
$$ 
$E$ is a Lie algebroid  over $A$ of rank $2n$, with vector bundle projection $\tau_{A}^{1}:E\ra A$, the projection over the first factor; and  anchor map determined by $\rho_{\tau_{A}^1}:=pr_2:E\rightarrow TA$, the projection onto the second factor. The bracket of sections on $E$ is denoted by $\lcf\cdot,\cdot
\rcf_{\tau_A^{1}}$ (see \cite{Eduardo} for more details). $E$ can be also seen as a subset of $A\times A\times TA$ and rewritten as  
$$
E= \{(a,b,v_a)\in A\times A\times TA\mid \rho(b)=(T\tau_A)(v_a), v_a\in T_{a}A \hbox{ and }\tau_A(a)=\tau_A(b)\}.
$$ 
Therefore, if $(a,b,v_a)\in E$, then $$\rho_{\tau_{A}^1}(a,b,v_{a})=(a,v_a)\in T_{a}A, \hbox{ and } \tau_A^{1}(a,b,v_a)=a\in A.$$

A local basis of sections of $E$, denoted by $\{\mathcal{X}_{\alpha},\mathcal{V}_{\alpha}\}$, is  given by
\begin{equation}\label{sectionsE}
\mathcal{X}_{\alpha}(e)=\left(e_{\alpha}(\tau_{A}(e)),\rho_{\alpha}^{i}\frac{\partial}{\partial x^{i}}\Big{|}_{e}\right),\quad \mathcal{V}_{\alpha}(e)=\left(0,\frac{\partial}{\partial y^{\alpha}}\Big{|}_{e}\right),
\end{equation}
for $e\in(\tau_{A})^{-1}(U).$  The former basis of sections  induces local coordinates $(x^{i},y^{\alpha};z^{\alpha},v^{\alpha})$ on  $E$ and  it determines the local structure of $(E,\lcf\cdot,\cdot\rcf_{\tau_{A}^{1}},\rho_{\tau_{A}^{1}})$ (see \cite{Eduardo} for further details).

Consider  the vector bundle projection $\tau_{A^{*}}:A^{*}\rightarrow M$ of the dual vector bundle $A^{*}$ of $A$ and the prolongation $\mathcal{T}^{\tau_{A^{*}}}A$ of $A$ over $\tau_{A^{*}}$ which is defined by
$$
E^{*}:=\mathcal{T}^{\tau_{A^{*}}}A=\{(b,v)\in A\times TA^{*}\mid \rho(b)=(T\tau_{A^{*}})(v)\},
$$
Alternatively $E^{*}$ can be seen as a subset of $A^{*}\times A\times TA^{*}$ given by 
\begin{align*}E^*=\{(a^{*},b,v_{a^{*}})\in A^*\times A\times TA^{*}&\mid \rho(b)=(T\tau_{A^{*}})(v_{a^{*}}),\\
&\qquad\qquad v_{a^{*}}\in T_{a}A{^*}, \tau_{A^{*}}(a^{*})=\tau_{A}(b)\}.\end{align*}

The prolongation $E^{*}$ is a Lie algebroid over $A^{*}$ of rank $2n$, with canonical projection denoted by $\tau_{A^{*}}^{1}:=pr_1:E^{*}\ra A^*$,  given the projection over the first factor as bundle projection and anchor map denoted by $\rho_{\tau_{A^{*}}^{1}}:E^*\rightarrow TA^{*}$.

If $\{e^{\alpha}\}$ is the dual basis of $\{e_{\alpha}\}$, then $\{{\mathcal Y}_{\alpha},{\mathcal
P}^{\alpha}\}$ is a basis of sections of the vector bundle $E^{*}$, where
\begin{equation}\label{sectionsEs}
{\mathcal Y}_{\alpha}(e^*)=\left(e_{\alpha}(\tau_{A^{*}}(e^*)),\rho_{\alpha}^i\displaystyle\frac{\partial
}{\partial x^i}\Big{|}_{e^*}\right),\quad
{\mathcal P}^{\alpha}(e^*)=\left(0,\displaystyle\frac{\partial }{\partial
p_{\alpha}}\Big{|}_{e^*}\right),
\end{equation}
for $e^*\in (\tau_{A^{*}})^{-1}(U).$ Here, $(x^i,p_{\alpha})$ are local coordinates on $A^*$ induced by the local coordinates $(x^i)$ and the basis $\{e^{\alpha}\}$. Using the basis $\{{\mathcal Y}_{\alpha},{\mathcal P}^{\alpha}\}$, it is possible to induce local coordinates $(x^i,y^{\alpha}; p_{\alpha}, \bar{p}_{\alpha})$ on $E^*$ . If $\omega^*$ is a point of  $(\tau_{A^{*}}^{1})^{-1}((\tau_{A^{*}})^{-1}(U))$, then $(x^i,p_{\alpha})$ are the coordinates of the point $\tau_{A^{*}}^{1}(\omega^*)\in \tau_{A^{*}}^{-1}(U)$ and
$$
\omega^*=y^{\alpha}{\mathcal Y}_{\alpha}(\tau_{A^{*}}^{1}(\omega^*)) + \bar{p}_{\alpha} {\mathcal
P}^{\alpha}(\tau_{A^{*}}^{1}(\omega^*)).
$$
Besides that, this basis allows us to provide the local structure of the Lie algebroid $E^{*}$, 
(see \cite{LMM} for the derivation of local expressions of the bracket and anchor map on $E^{*}$). 

\subsection{Symplectic Lie algebroids and Lie subalgebroids}
In this subsection we regard some results given in \cite{LMM} on symplectic Lie algebroids and Lie subalgebroids.

\subsubsection{The canonical symplectic structure of the Lie algebroid $E^{*}$}

Consider $\lambda_{A}$ the \textit{Liouville section} of $E^{*}$, which is defined in such way that, if $a^{*}\in A^{*}$ and $(b,v)$ is a point of the fiber of $E^{*}$
over $a^{*}$, then $\displaystyle{\lambda_{A}(a^{*})(b,v)=a^{*}(b)}$.

The $2$-section on $E^*$ defined by
$\displaystyle{\Omega_{A}:=-d^{E^{*}}\lambda_{A}}$,  is non-degenerate satisfying $d^{E^{*}}\Omega_{A}=0$, that is, $\displaystyle{\Omega_{A}}$ is a symplectic section of the Lie algebroid
$E^{*}$,  called \textit{canonical symplectic section} (see \cite{LMM} for instance). In local coordinates, $\Omega_{A}=\mathcal{Y}^{\alpha}\wedge\mathcal{P}_{\alpha}+\frac{1}{2}\mathcal{C}_{\alpha\beta}^{\gamma}\mathcal{Y}^{\alpha}\wedge\mathcal{Y}^{\beta}$, where $\{{\mathcal Y}^{\alpha},{\mathcal
P}_{\alpha}\}$ is the dual basis of the basis $\{{\mathcal Y}_{\alpha},{\mathcal P}^{\alpha}\}$ and $\mathcal{C}_{\alpha\beta}^{\gamma}$ denotes the constant structures of the  Lie bracket$\lcf\cdot,\cdot
\rcf$ with respect to the basis $\{e_{\alpha}\}$.

 \subsubsection{Symplectic Lie algebroid and Lagrangian Lie subalgebroids}

Next, we review the notion of symplectic Lie algebroid following \cite{LMM}:

\begin{definition}
A Lie algebroid $(A,\lcf\cdot,\cdot\rcf,\rho)$ over a manifold $M$
is said to be symplectic if  admits a symplectic section $\Omega,$
that is, a section of the vector bundle
$\bigwedge^{2}A^{*}\ra M$, such that:
\begin{itemize}
\item[(i)] For all $x\in M,$ the 2-form $\Omega_{x}:A_{x}\times
A_{x}\ra\R$ in the vector space $A_{x}$ is nondegenerate; and
\item[(ii)] $\Omega$ is a 2-cocycle, that is, $d^{A}\Omega=0$, with $d^{A}$ denoting the differential of $A$.
\end{itemize}
\end{definition}

A notable example of a symplectic Lie algebroid is $(E^{*},\lcf\cdot,\cdot\rcf_{\tau_{A^{*}}^{1}},\rho_{\tau_{A^{*}}^{1}})$  with the canonical symplectic section $\Omega_{A}$.

We recall 
that if $A$ is a symplectic Lie algebroid
with symplectic section $\Omega$ then the prolongation
$E$ of $A$ over the vector bundle projection
$\tau_A:A\to M$ is a symplectic Lie algebroid with symplectic
section determined by the complete lift of $\Omega.$

\begin{remark}
If $(M,\Omega_{TM})$ is a symplectic manifold, then
 $E^{*}=T(T^{*}M)$ 
 is a Lie algebroid over $A=TM$. In this case, the Liouville section and the canonical symplectic section of $E^{*}$ are, respectively, 
$\lambda_{A}=\lambda_{TM}$ the usual Liouville $1-$form on
$T^{*}M$ and $\Omega_{A}=\Omega_{TM}$ the canonical symplectic
$2-$form on the cotangent bundle $T^{*}M.$ Moreover, the tangent
bundle of $M$ is a symplectic Lie algebroid and the complete lift
$\Omega_{TM}^{\bf c}$ of $\Omega_{TM}$ to $TM$ is a symplectic
2-form on $TM.$
\end{remark}

\subsubsection{Lagrangian Lie subalgebroids}
Now, we are interested in the notion of a Lagrangian Lie subalgebroid of a symplectic Lie algebroid.

\begin{definition}
Let $(A,\lcf\cdot,\cdot \rcf_A,\rho)$ be a Lie algebroid over a manifold $M$
and $N$ a submanifold of $M$. A Lie subalgebroid of $A$ over $N$
is a vector subbundle $F$ of $A$ over $N$, such that, $\rho_{F}=\rho\mid_{F}:F\to TN$ is well defined; and
given $X,Y\in\Gamma(F)$ and
$\widetilde{X},\widetilde{Y}\in\Gamma(A)$ arbitrary extensions of
$X,Y$, respectively, 
$(\lcf\widetilde{X},\widetilde{Y}\rcf_A)\mid_{N}\in\Gamma(F)$. 

The maps $i$ and $j$ denote the canonical inclusions of $F$ to $A$ and $N$ to $M$, respectively. We say that $j:F\to A$, $i:N\to M$ is a Lie subalgebroid of  the vector bundle projection $\tau_A:A\to M$.
\end{definition}

\begin{definition}\label{d7.2}
Let $(A,\lcf\cdot,\cdot\rcf,\rho)$ be a symplectic Lie algebroid
with symplectic section $\Omega$ 
 and $j:F\to A$, $i:N\to M$ a Lie
subalgebroid. The Lie subalgebroid is said to be Lagrangian if
$j(F_x)$ is a Lagrangian subspace of the symplectic vector space
$(A_{i(x)},\Omega_{i(x)}),$ for all $x\in N$. 
\end{definition}

Observe that definition \ref{d7.2} implies the following:
\begin{enumerate}
\item $\hbox{rank} F=\displaystyle\frac{1}{2}\hbox{rank} A$; and 
\item $(\Omega (i(x)))_{|j(F_x)\times j(F_x)}=0,$ for all $x\in N.$
\end{enumerate}

\begin{remark}\label{r7.3}
If $M$  is a symplectic manifold, $S$ a submanifold of
$M$ and $i: S \to M$ the canonical inclusion, then the standard
Lie algebroid $\tau_{TM}:TM\to M$ is symplectic and $i: S \to M$, $j
= Ti: TS \to TM$ is a Lie subalgebroid of $\tau_{TM}: TM \to M$.
Furthermore, $S$ is a Lagrangian submanifold of $M$ if and only if the
Lie subalgebroid $i:S\to M, j=Ti:TS\to TM$ of $\tau_M:TM\to M$ is
Lagrangian.
\end{remark}

\subsection{Admissible elements}

A curve $\gamma:I\subset\R\ra A$ is said to be \textit{admissible} if
$(\gamma(t),\dot{\gamma}(t))\in E_{\gamma(t)}$, for all $t\in I$. 
Locally, we have that 
$\gamma(t)=(x^{i}(t),y^{\alpha}(t))$ is an admissible curve if and only if 
$$
\frac{dx^{i}}{dt}=\rho_{\alpha}^{i}y^{\alpha}, \quad \hbox{for } i\in\{1,\ldots,m\}.
$$
In other words, $\gamma:I\subset\R\ra A$ is said to be admissible if and only if $\displaystyle{\frac{d}{dt}(\tau_{A}\circ\gamma)=\rho\circ\gamma}$.

\begin{definition}
Let $(A,\lcf\cdot,\cdot \rcf_A,\rho)$ be a Lie algebroid over a manifold $M$, with $\tau_A:A\to M$ the vector bundle projection. A tangent vector $v$ to $A$ at the point $a\in A$ is called \textit{admissible} if $\rho(a)=T_{a}\tau_A(v)$. In addition, a curve in $A$ is admissible if its tangent vectors are admissible. The set of admissible tangent vectors to $A$ is represented by $A^{(2)}$.
\end{definition}

Notice that $A^{(2)}$ is a subset of  the prolongation $E$ of $A$ over $\tau_A$ which can be written as 
$$
A^{(2)}=\{(a,v_a)\in A\times TA\mid \rho(a)=T\tau_A (v_a)\} \subset E.
$$

Given $(a,b,v_a)\in E$ such that $\rho(a)=T\tau_A (v_a)$, we have the two canonical inclusions
\begin{eqnarray*}
i_{TA}:A^{(2)}&\hookrightarrow& TA,\quad\qquad\qquad i_{E}:A^{(2)}\hookrightarrow E\\
 (a,v_a)&\longmapsto& v_a\qquad\quad\qquad\qquad (a,v_a)\longmapsto (a,a,v_a).
\end{eqnarray*}

In the context of our work, $A^{(2)}$ plays the role  of the second order tangent bundle used in classical mechanics \cite{Eduardo}. In local coordinates, the set $A^{(2)}$ is characterized by the condition 
$$
\{(x^{i},y^{\alpha};z^{\alpha},v^{\alpha})\in E\mid y^{\alpha}=z^{\alpha}\}
$$ and hence $(x^i,y^{\alpha},v^{\alpha})$ are considered local coordinates on $A^{(2)}$.

\section{Lagrangian submanifolds generating second-order dynamics}

\subsection{The prolongation of a prolongated Lie algebroid}

Given a Lie algebroid $A$ and the $A$-tangent bundle $E$, we construct  the prolongation of $E$ over $\tau_{A}^{1}:E\to A$, a Lie algebroid over $E$ denoted by $\mathcal{T}^{\tau_{A}^{1}}E$ and with $\tau_{A}^{2}:\mathcal{T}^{\tau_A^{1}}E\ra E$ representing the bundle projection. Since the prolongation over a fiber map of a Lie algebroid is a Lie algebroid, $\mathcal{T}^{\tau_{A}^{1}}E$ is  indeed a Lie algebroid. 
We have
\begin{align}\label{eqq1}
\mathcal{T}^{\tau_{A}^{1}}E=&\Big{\{}(z,y,\chi)\in E\times E\times TE\mid\tau_{A}^{1}(z)=\tau_{A}^{1}(y),\chi\in T_{z}E,\,\rho_{\tau_{A}^{1}}(y)=T_{z}\tau_{A}^{1}(\chi)\Big{\}}\nonumber\\
=&\Big{\{}(z,\chi)\in E\times TE\mid\rho_{\tau_{A}^{1}}(z)=T\tau_{A}^{1}(\chi)\Big{\}}.
\end{align}

As we refer before, a basis of sections of $A$ brings a local basis  of sections  of $E$ given by \eqref{sectionsE} which induces local coordinates $(x^{i},y^{\alpha};z^{\alpha},v^{\alpha})$ on $E$. From this basis, we construct a local basis of sections of $\mathcal{T}^{\tau_{A}^{1}}E$ as follows: consider  $s=(a,v_b)\in E$ and define the element of the basis $\{\mathcal{X}_{\alpha}^{1},\mathcal{X}_{\alpha}^{2},\mathcal{V}_{\alpha}^{1},\mathcal{V}_{\alpha}^{2}\}$ as
\begin{align*}\label{sectionsTE}
\mathcal{X}_{\alpha}^{1}(s)&=\left(\mathcal{X}_{\alpha}(b),\rho_{\alpha}^{i}\frac{\partial}{\partial x^{i}}\Big{|}_{s}\right), 
\mathcal{X}_{\alpha}^{2}(s)=\left(0,\frac{\partial}{\partial z^{\alpha}}\Big{|}_{s}\right), \\
\mathcal{V}_{\alpha}^{1}(s)&=\left(\mathcal{V}_{\alpha}(b),\frac{\partial}{\partial y^{\alpha}}\Big{|}_{s}\right), \quad
\mathcal{V}_{\alpha}^{2}(s)=\left(0,\frac{\partial}{\partial u^{\alpha}}\Big{|}_{s}\right).
\end{align*} 
The Lie algebroid structure of $(\mathcal{T}^{\tau_{A}^{1}}E,\lcf\cdot,\cdot\rcf_{\tau_{{A}}^{2}},\rho_{\tau_{A}^{2}})$ is characterized by
$$
\rho_{\tau_{A}^{2}}({\mathcal X}_{\alpha}^{1})=\rho_{\alpha}^i\displaystyle\frac{\partial
}{\partial x^i},\quad\rho_{\tau_{A}^{2}}({\mathcal
X}_{\alpha}^{2})=\displaystyle\frac{\partial
}{\partial z^{\alpha}},\quad
\rho_{\tau_{A}^{2}}(\mathcal{V}_{\alpha}^{1})=\displaystyle\frac{\partial}{\partial y^{\alpha}},\quad\rho_{\tau_{A}^{2}}(\mathcal{V}_{\alpha}^{2})=\displaystyle\frac{\partial}{\partial u^{\alpha}},
$$ 
and the Lie bracket of sections, where the unique non-zero Lie brackets of sections are $\lcf{\mathcal X}_{\alpha}^{1},{\mathcal X}_{\beta}^{1}\rcf_{\tau_{A}^{2}}={\mathcal C}_{\alpha\beta}^{\gamma}{\mathcal X}_{\gamma}^{1}$, for all $\alpha,\beta$ and $\gamma$. Such a basis induces local coordinates $(x^{i},y^{\alpha},z^{\alpha},v^{\alpha};e^{\alpha};d^{\alpha},w^{\alpha},\bar{z}^{\alpha})$ on $\mathcal{T}^{\tau_A^{1}}E.$

Consider 
\begin{equation}\label{eqq2}
\mathcal{T}^{\tau_{A}^{1}}E^{*}=\Big{\{}(w^{*},z_{w})\in E^{*}\times TE\mid\rho_{\tau_{A^{*}}^{1}}(w^{*})=T\tau_{A}^{1}(z_{w})\Big{\}}.
\end{equation} Similarly, from the basis of sections of $E^{*}$ given by \eqref{sectionsEs}, we define the basis  of sections of $\mathcal{T}^{\tau_{A}^{1}}E^{*}$, denoted $\{\mathcal{Y}_{\alpha}^{1},\mathcal{Y}_{\alpha}^{2},(\mathcal{P}^{\alpha})^{1},(\mathcal{P}^{\alpha})^{2}\}$,
as follows:
\begin{align*}
\mathcal{Y}_{\alpha}^{1}(\bar{s})&=\left(\mathcal{Y}_{\alpha}(b^{*}),\rho_{\alpha}^{i}\frac{\partial}{\partial x^{i}}\Big{|}_{\bar{s}}\right),\quad
\mathcal{Y}_{\alpha}^{2}(\bar{s})=\left(0,\frac{\partial}{\partial p_{\alpha}}\Big{|}_{\bar{s}}\right),\\
(\mathcal{P}^{\alpha})^{1}(\bar{s})&=\left(\mathcal{P}^{\alpha}(b^{*}),\frac{\partial}{\partial y^{\alpha}}\Big{|}_{\bar{s}}\right),\quad
(\mathcal{P}^{\alpha})^{2}(\bar{s})=\left(0,\frac{\partial}{\partial \bar{p}_{\alpha}}\Big{|}_{\bar{s}}\right),
\end{align*}
where $\bar{s}=(a,v_{b^{*}})\in E^{*}$
and $(x^{i},y^{\alpha};p_{\alpha},\bar{p}_{\alpha})$ are local coordinates on $E^{*}$ induced by the its basis of sections. It follows that the Lie algebroid structure of $(\mathcal{T}^{\tau_{A}^{1}}E^{*}; \lcf\cdot,\cdot\rcf_{\tau_{{A}^{*}}^{2}} \rho_{\tau_{{A}^{*}}^{2}})$ is determined by the unique non-zero Lie bracket  $\lcf{\mathcal Y}_{\alpha}^{1},{\mathcal Y}_{\alpha}^{2}\rcf_{\tau_{{A}^{*}}^{2}} =\mathcal{C}_{\alpha\beta}^{\gamma}\mathcal{Y}_{\gamma}^{1}$ and 
\begin{align*}
\rho_{\tau_{{A}^{*}}^{2}}(\mathcal{Y}_{\alpha}^{1})=&\rho_{\alpha}^{i}\frac{\partial}{\partial x^{i}},\quad\rho_{\tau_{{A}^{*}}^{2}}(\mathcal{Y}_{\alpha}^{2})=\frac{\partial}{\partial p^{\alpha}},\\
\rho_{\tau_{{A}^{*}}^{2}}((\mathcal{P}^{\alpha})^{1})=&\frac{\partial}{\partial y^{\alpha}},\quad \rho_{\tau_{{A}^{*}}^{2}}((\mathcal{P}^{\alpha})^{2})=\frac{\partial}{\partial \bar{p}_{\alpha}}.\end{align*}

Moreover, local coordinates $(x^{i},y^{\alpha},p_{\alpha},\bar{p}_{\alpha}; q^{\alpha}, \bar{q}^{\alpha}; l_{\alpha},\bar{l}_{\alpha})$ on $\mathcal{T}^{\tau_{A}^{1}}E^{*}$ are  induced by the above basis.

Next, we introduce a result which allows us to consider a Lie algebroid that plays a fundamental role  in the framework of this work (see \cite{Mac} for the proof).
\begin{lemma}\label{th1}
Consider   the Lie algebroid $E=\mathcal{T}^{\tau_A}A$ over $A$ with projection $\tau_A^{1}:E\rightarrow A$ and  the Lie algebroid $\mathcal{T}^{\tau_A^{1}}E$ over $E$. If $N$ is a submanifold of $E$ with inclusion $i_{N}:N\to E$ and the anchor map $\rho_{\tau_{A}^{2}}:\mathcal{T}^{\tau_A^{1}}E\rightarrow TE$ has constant rank, then the vector bundle $A_N=\rho_{\tau_{A}^{2}}^{-1}(TN)$ determined by $\tau_{A}^{2}{\big{|}_{\rho_{\tau_{A}^{2}}^{-1}(TN)}}:A_N\rightarrow N$ is a Lie subalgebroid over $N$, where $\tau_{A}^{2}:\mathcal{T}^{\tau_A^{1}}E\ra E$ is the bundle projection.
\end{lemma}

\begin{remark}
Observe that when $A=TM,$ the prolongation $E=TTM$ is a Lie algebroid over $TM$ with projection $\tau_{TM}:TTM\rightarrow TM$ and $T(TTM)$ is a Lie algebroid over $TTM$ with canonical projection $\tau_{TA}:TE\to TA$. If $N$ is a submanifold of $TTM$, then $TN$ is a Lie subalgebroid of $TE$ over $N$. Hence, for $N=T^{(2)}M$, the second-order tangent bundle of a manifold $M$, we get that $A_N=T(T^{(2)}M)$ is a Lie subalgebroid of $T(TTM)$ over $T^{(2)}M$. 

\end{remark} 

Consider $N$ as being the set of admissible elements of $A$, $N=A^{(2)}$, locally determined by the induced coordinates $(x^i,y^{\alpha}, v^{\alpha})$, where $(x^i,y^{\alpha}; z^{\alpha}, v^{\alpha})$ are local coordinates on $E$ and $\rho_{\tau_{A}^{1}}(x^i,y^{\alpha}; z^{\alpha}, v^{\alpha})=(x^i,z^{\alpha},\rho_{\alpha}^{i}y^{\alpha},v^{\alpha})$. Therefore, the  induced coordinates on the Lie subalgebroid $\rho_{\tau_{A}^{2}}^{-1}(TA^{(2)})$ are given by $(x^{i}, z^{\alpha}, v^{\alpha},\rho_{\alpha}^{i}y^{\alpha},\bar{u}^{\alpha},w^{\alpha})$. 

\subsection{Tulczyjew's isomorphisms}
Here we present two vector bundle isomorphisms with the aim of extend  \textit{Tulczyjew's triple} proposed in \cite{LMM} to a prolongation Lie algebroid as a generalization of  Tulczyjew's  symplectomorphisms \cite{Tulczy1}. For this,  we just need to specify the construction of \cite{LMM} to our situation, since $E$ is a Lie algebroid and with that  we achieve an important step in our construction.

The vector bundle isomorphisms to consider,
$\alpha_{E}:\mathcal{T}^{\tau_{A^{*}}^{1}}E\to
(\mathcal{T}^{\tau_{A^{*}}^{1}}E)^{*}$ and
$\beta_{E^{*}}:\mathcal{T}^{\tau_{A^{*}}^{1}}E\rightarrow (\mathcal{T}^{\tau^1_{A^{*}}}E)^{*}$, are defined in such a way the following diagram is commutative, 
\begin{equation}\label{Diag}
\begin{picture}(200,83)(75,20)
\put(125,40){\makebox(0,-1){$E$}}
\put(155,70){\vector(-3,-2){30}}
 \put(240,70){\vector(-3,-2){30}}
\put(210,40){\makebox(0,-1){$ E^*$}}
\put(65,60){$(\tau_{A}^{2})^{*}$}
\put(160,55){$\quad\tau_{A^{*}}^{2}$}
\put(235,60){$\quad(\tau_{{A}^{*}}^{2})^*$}\put(90,70){\vector(3,-2){30}}
\put(140,50){$\hbox{pr}_{1}$} \put(175,70){\vector(3,-2){30}}
\put(60,80){\makebox(0,0){$(\mathcal{T}^{\tau_A^{1}}E)^{*}$}}
\put(110,90){$\alpha_{E}$}\put(200,90){$\beta_{E^{*}}$}\put(140,80){\vector(-3,0){50}}
\put(160,80){\makebox(0,0){$\mathcal{T}^{\tau_{A^{*}}^{1}}E$}}
\put(185,80){\vector(3,0){50}}
\put(275,80){\makebox(0,0){$(\mathcal{T}^{\tau_{A^{*}}^{1}}E)^{*}$}}\end{picture}
\end{equation}

The vector bundle isomorphism $\beta_{E^{*}}$ is determined by $$\beta_{E^{*}}(v,z)=i(v,z)(\Omega_{E}(\tau_{A^*}^{2}(v,z))),$$ for $(v,z)\in\mathcal{T}^{\tau_{A^*}^{1}}E$ with $v=(b,v_a)\in E$ and $z\in T_{v}E^{*}$, being $\Omega_{E}$  the canonical symplectic section of  $\mathcal{T}^{\tau_{A^*}^{1}}E.$

In order to define $\alpha_E$, we need to introduce first a canonical involution. Let $(v,w)$ be a point in $(\mathcal{T}^{\tau_{A}^{1}}E)_e$ with $e\in E_x$, $x\in A$,  that is, $\tau^{2}_{A}(v,w)=e$ such that $x=\tau_A^{1}(e)=\tau_{A}^{1}(v)$. Hence, there exists one and only one tangent vector $\tilde{v}\in T_{v}E$ such that $\tilde{v}(f\circ\tau^{1}_A)=(d^{E}f)(x)(e)$ for $f\in C^{\infty}(A)$ and thus  $(e,\tilde{v})$ is an element of $(\mathcal{T}^{\tau_A^{1}}E)_{v}$. Indeed,  the vector $\tilde{v}$ projects to $\rho_{\tau_{A}^{1}}(e)$, since $((T_{v}\tau_{A}^1)(\tilde{v}))f=\rho_{\tau_{A}^{1}}(e)f$ for all $f\in C^{\infty}(A).$

\begin{definition}
The \textit{canonical involution} associated with the Lie algebroid $A$ is  the smooth involution $\sigma_{E}:\mathcal{T}^{\tau_A^{1}}E\rightarrow\mathcal{T}^{\tau_{A}^{1}}E$ given by $\sigma_{E}(v,w)=(e,\tilde{v})$.
\end{definition}
Notice that, the map $\sigma_{E}$ satisfies $\sigma_{E}^2=Id$ and $\bar{pr} \circ \sigma_{E}= \tau^{2}_A$, where $\bar{pr} :\mathcal{T}^{\tau_A^{1}}E\to E$ is the canonical projection onto $E$. 

The  canonical involution allows us to define the vector bundles isomorphism $\alpha_E:\mathcal{T}^{\tau_{A^*}^{1}}E\rightarrow (\mathcal{T}^{\tau_A^{1}}E)^{*}$ as follows: let $\langle\cdot,\cdot\rangle:E\times E^{*}\to\mathbb{R}$ be the pairing given by $\langle w,w^{*}\rangle=w^{*}(w),$ for $w\in E_x$ and $w^{*}\in E^{*}_x$, with $x\in A.$ If $v\in E$, $(v,z_w)\in ({\mathcal T}^{\tau_{A}^{1}}E)^{*}_v$ and $(v,z_{w^{*}})\in ({\mathcal T}^{\tau_{A^*}^{1}}E)_v$, then $
(z_w,z_{w^{*}})\in T_{(w,w^{*})}(E\times E^{*})$ where 
$$
T_{(w,w^{*})}(E\times E^{*})=\{(z'_w,z'_{w^{*}})\in T_{w}E \times T_{w^*}E^{*}\mid(T_w \tau_{A}^{1})(z'_w)=(T_{w^{*}}\tau_{A^{*}}^{1})(z'_{w^{*}})\}.
$$

We may now consider the map $\widetilde{T\langle\cdot,\cdot\rangle}:({\mathcal T}^{\tau_{A}^{1}}E)^{*}\times{\mathcal T}^{\tau_{A^*}^{1}} E\rightarrow\R$ defined by
$$
\tpair((v,z_w);(v,z_{w^{*}}))=dt_{\langle w,w^*\rangle}((T_{(w,w^*)}\langle\cdot,\cdot\rangle)(z_w,z_{w^*})),
$$ 
where  $t\in\R$ and $T\langle\cdot,\cdot\rangle:T(E\times E^{*})\ra T\R$ is the tangent map to the pairing. The pairing induces an isomorphism between the vector bundles $\mathcal{T}^{\tau_A^{1}}E\to E$ and $(\mathcal{T}^{\tau_{A^{*}}^{1}}E)^*\to E$ which we also  denote by $\widetilde{T{\langle\cdot,\cdot\rangle}}$, that is, $\widetilde{T{\langle\cdot,\cdot\rangle}}:\mathcal{T}^{\tau_A^{1}}E\to (\mathcal{T}^{\tau_{A^{*}}^{1}}E)^*$, because this is a non-singular pairing (see \cite{LMM}).

Hence, we can consider the isomorphism of vector bundles $\alpha^*_{E}:{\mathcal T}^{\tau_A^{1}}E\to (\mathcal{T}^{\tau_{A^{*}}^{1}}E)^{*}$ given by
$$
 \alpha_{E}^*=\widetilde{T{<\cdot,\cdot>}}\circ \sigma_{E},
 $$
 with $\sigma_{E}:{\mathcal T}^{\tau_A^{1}} E\to{\mathcal T}^{\tau_A^{1}} E$ the canonical involution. Finally, the isomorphism $\alpha_{E}:\mathcal{T}^{\tau_{A^{*}}^{1}}E\to ({\mathcal T}^{\tau_A^{1}}E)^*$ between the vector bundles $\mathcal{T}^{\tau_{A^{*}}^{1}}E\to E$ and $({\mathcal T}^{\tau_A^{1}} E)^*\to E$ is considered simply as the dual map to $\alpha_{E}^*:{\mathcal T}^{\tau_A^{1}}E \to (\mathcal{T}^{\tau_{A^{*}}^{1}}E)^*.$ The diagram (\ref{Diag}) is designed by \textit{Tulczyjew's triple for the prolongated Lie algebroid} $\mathcal{T}^{\tau_{A}^{1}}E$.

By considering adapted coordinates  $(x^{i},y^{\alpha},p_{\alpha},\bar{p}_{\alpha}; q^{\alpha},\bar{q}^{\alpha}, l_{\alpha},\bar{l}_{\alpha})$ from the basis of sections $\{\mathcal{Y}_{\alpha}^{(1)},\mathcal{Y}_{\alpha}^{(2)},(\mathcal{P}^{\alpha})^{(1)},(\mathcal{P}^{\alpha})^{(2)}\}$ of $\mathcal{T}^{\tau_{A}^{1}}E^{*}$, the local representation of $\alpha_{E}$ is given by
\begin{equation}\label{triple}
\alpha_{E}(x^{i},y^{\alpha},p_{\alpha},\bar{p}_{\alpha}; q^{\alpha},\bar{q}^{\alpha}, 
l_{\alpha},\bar{l}_{\alpha})=(x^{i},y^{\alpha},q^{\alpha},\bar{q}^{\alpha};
l_{\alpha}+\mathcal{C}_{\alpha\beta}^{\gamma}p_{\gamma}q^{\beta},\bar{l}_{\alpha},p_{\alpha},\bar{p}_{\alpha}).
\end{equation}

\begin{remark}
If $A=TM,$ then $E=TTM$ and it is straightforward to note that the vector bundle isomorphisms $\alpha_{TTM}:T(T^{*}TM)\ra T^{*}(TTM)$ and $\beta_{(TTM)^{*}}:T(T^{*}TM)\ra T^{*}(T^{*}TM)$ are the ones constructed by  Le\'on and Lacomba in \cite{LeonLacomba}.
\end{remark}

\subsection{Lagrangian submanifolds generating second-order Lagrangian dynamics}\label{section3.3}
Recall that given a finite-dimensional symplectic manifold $(M,\omega)$ and a submanifold $N$ of $M$, with canonical inclusion $i_N: N\hookrightarrow M$, then $N$ is a \textit{Lagrangian submanifold} if $i^{*}_{N}\,\omega=0$ and $\mbox{dim}\hspace{1mm}N=\frac{1}{2}\mbox{dim}\hspace{1mm}M$.

It is well know that given a manifold $M$ and a function $S:M\to\R$, the submanifold ${\rm Im }\;  \hbox{d}S\subset T^{*}M$ is Lagrangian \cite{weinstein}. There is a more general construction of the authors \'Sniatycki and Tulczyjew \cite{ST} (see also \cite{Tulczy1} and \cite{art:Tulczyjew76_1}) which we use to generate the dynamics.

\begin{theorem}[\'Sniatycki and Tulczyjew \cite{ST}]\label{tulczyjew}
Let $M$ be a smooth manifold, $N \subset M $ a submanifold, and $S\colon N \rightarrow {\mathbb R}$.  Then,
\begin{align*}
\Sigma _S = \bigl\{ \mu \in &T ^\ast M \mid \pi _M (\mu) \in N \text{ and }\\ 
\left\langle \mu, v \right\rangle &= \left\langle \mathrm{d} S , v, \right\rangle \hbox{ for all } v \in T N \subset T M \hbox{ such that } \tau  _M (v) = \pi _M (\mu) \bigr\}
\end{align*}
is a Lagrangian submanifold of $ T ^\ast M$, where $\pi_M:T^{*}M\to M$ and $\tau_M:TM\to M$ denote the cotangent and tangent bundle projections, respectively.
\end{theorem}

Turning back to the Lie algebroid formulation,  consider a Lagrangian function $L:A^{(2)}\rightarrow\R$ and the following Lagrangian Lie subalgebroid of $(\mathcal{T}^{\tau_A^{1}}E)^{*}$ (see Appendix A) 
$$
\Sigma_{A_N}=\{\mu\in(\mathcal{T}^{\tau_A^{1}}E)^{*} \mid i^{*}_{A_N}\mu=d^{A_N}L\},
$$ where $A_N=\rho_{\tau_{A}^{2}}^{-1}(TN),$ $i_{A_N}:A_N\rightarrow\mathcal{T}^{\tau_A^{1}}E$ is the canonical inclusion and $N=A^{(2)}.$ Here $d^{A_N}$ denotes the differential operator of the Lie subalgebroid $A_N$ over $A^{(2)}$.

Consider induced coordinates $(x^{i},y^{\alpha}; z^{\alpha},v^{\alpha})$ on $E$.  As we commented before, the set of admissible elements is characterized by the condition $y^{\alpha}=z^{\alpha}$ and induced coordinates on $A^{(2)}$ are $(x^{i},y^{\alpha},v^{\alpha})$. Locally, the submanifold $\Sigma_{A_N}$ is characterized by
\begin{align}
\mu_{\alpha}&=\rho_{\alpha}^{i}\frac{\partial L}{\partial x^{i}}\label{eqmu1}\\
\bar{\mu}_{\alpha}+\check{\mu}_{\alpha}&=\frac{\partial L}{\partial y^{\alpha}}\label{eqmu2}\\
\widetilde{\mu}_{\alpha}&=\frac{\partial L}{\partial v^{\alpha}}\label{eqmu3}
\end{align}
where
$\{\mu_{\alpha},\bar{\mu}_{\alpha},\check{\mu}_{\alpha},\widetilde{\mu}_{\alpha}\}$ are sections of $(\mathcal{T}^{\tau_A^{1}}E)^{*}.$

The following result gives rise to a special Lagrangian submanifold. Such a Lie submanifold generates the dynamics for systems defined on $A^{(2)}$ and it is a result consequence of Corollary $8.3$ in \cite{LMM}.
\begin{proposition}
$S_L=\alpha^{-1}_{E}(\Sigma_{A_N})=\{x\in\mathcal{T}^{\tau_{A^{*}}^{1}}E\mid\alpha_{E}(x)\in\Sigma_{A_N}\}\subset\mathcal{T}^{\tau_{A^{*}}^{1}}E$ is a Lagrangian submanifold of $\mathcal{T}^{\tau_{A^{*}}^{1}}E.$
\end{proposition}

Given a Lagrangian $L$, the Lagrangian submanifold $S_L$  determines an implicit differential equation on $\mathcal{T}^{\tau_{A^*}^{1}}E$. A curve $\gamma$ in $S_L$, which maps $t\in\R\mapsto(\gamma_{1}(t),\gamma_2(t))\in S_L$, is called a \textit{solution of the Lagrangian problem} determined by $L:A^{(2)}\ra\R$ if the curve $\gamma_2:I\ra TE^{*},$ $t\mapsto \gamma_{2}(t)$ is a tangent lift; that is, $\gamma_{2}(t)=\dot{c}(t),$ where $c:I\ra E^{*}$ is a curve in $A^{*}$ given by $T\tau_{A^{*}}^{1}\circ\gamma_2,$ with $T\tau_{A^{*}}^{1}:TE^{*}\ra E^{*}$ the canonical projection.

To obtain solutions of the dynamics given by $S_L$ 
it is needed to extract the integrable part from a sequence of submanifolds of $\mathcal{T}^{\tau_{A^{*}}^{1}}E;$ that is, if $\tau_{A^{*}}^{2}:\mathcal{T}^{\tau_{A^{*}}^{1}}E\rightarrow E^*$ then the algorithm is given by (see Remark $3.1$ in \cite{IMMS} and Appendix B): 
$$
S^{0}:=S_L,\quad S^{1}=S^0\cap\rho_{\tau_{A}^{1}}^{-1}(T(\tau_{A^{*}}^{2}(S^{0}))), \ldots, S^{(l+1)}=S^{l}\cap\rho_{\tau_{A}^{1}}^{-1}(T(\tau_{A^{*}}^{2}(S^{l}))),\ldots
$$

When the algorithm stabilizes, one obtains the second-order Euler-Lagrange equations on Lie algebroids. 

Denoting by $(x^{i},y^{\alpha},p_{\alpha},\bar{p}_{\alpha}; z^{\alpha},v^{\alpha}; l_{\alpha},\bar{l}_{\alpha})$  induced coordinates by the basis of sections of $\mathcal{T}^{\tau_{A}^{1}}E^{*},$ $\{\mathcal{Y}_{\alpha}^{(1)},\mathcal{Y}_{\alpha}^{(2)},(\mathcal{P}^{\alpha})^{(1)},(\mathcal{P}^{\alpha})^{(2)}\}$ and using the expression of  Tulczyjew's isomorphism and the
condition $y^{\alpha}=z^{\alpha}$, equation \eqref{eqmu1}, \eqref{eqmu2} and \eqref{eqmu3} are equivalents to 
\begin{align} 
l_{\alpha}+\mathcal{C}_{\alpha\beta}^{\gamma}p_{\gamma}z^{\beta}&=\rho_{\alpha}^{i}\frac{\partial L}{\partial x^{i}},\label{eqmu4}\\ 
\bar{l}_{\alpha}+p_{\alpha}&=\frac{\partial L}{\partial y^{\alpha}},\label{eqmu5}\\
\bar{p}_{\alpha}&=\frac{\partial L}{\partial v^{\alpha}}.\label{eqmu6}
\end{align}
Differentiating with respect to the time equation \eqref{eqmu6}, using the condition $\dot{\bar{p}}_{\alpha}=\bar{l}_{\alpha}$ and replacing into \eqref{eqmu5} we have that
\begin{equation}\label{eqp}
p_{\alpha}=\frac{\partial
L}{\partial y^{\alpha}}-\frac{d}{dt}\left(\frac{\partial L}{\partial
v^{\alpha}}\right).
\end{equation}

Differentiating with respect to time the last equation, using the condition $\dot{p}_{\alpha}=l_{\alpha}$ and replacing into \eqref{eqmu4} we obtain that
$$\frac{d}{dt}\left(\frac{\partial L}{\partial y^{\alpha}}\right)-\frac{d^{2}}{dt^{2}}\left(\frac{\partial L}{\partial v^{\alpha}}\right)+\mathcal{C}_{\alpha\beta}^{\gamma}p_{\gamma}z^{\beta}=\rho_{\alpha}^{i}\frac{\partial L}{\partial x^{i}}.$$

Finally, replacing in the last equation the equation \eqref{eqp}, we obtain that 
$$\frac{d^{2}}{dt^2}\left(\frac{\partial L}{\partial
v^{\alpha}}\right)-\mathcal{C}_{\alpha\beta}^{\gamma}z^{\beta}\frac{\partial L}{\partial y^{\alpha}}+\mathcal{C}_{\alpha\beta}^{\gamma}z^{\beta}\frac{d}{dt}\left(\frac{\partial L}{\partial v^{\alpha}}\right)+\rho_{\alpha}^{i}\frac{\partial L}{\partial x^{i}}-\frac{d}{dt}\left(\frac{\partial L}{\partial y^{\alpha}}\right)=0.$$

We summarize the previous derivation as follow:
\begin{proposition}
Let $A$ be a Lie algebroid over $M$ with bundle projection $\tau_{A}:A\ra M$,  $\mathcal{C}_{\alpha\beta}^{\gamma},$ $\rho_{\alpha}^{i}$ the  constant structures defined from the Lie algebroid structures of $A$ and $(x^{i},y^{\alpha},v^{\alpha})$local coordinates on $A^{(2)}$. The second-order Euler-Lagrange equation for the Lagrangian $L:A^{(2)}\ra\R;$ are given by 
\begin{align*}
\rho_{\alpha}^{i}y^{\alpha}&=\frac{dx^{i}}{dt},\quad  v^{\alpha}=\frac{dy^{\alpha}}{dt},\\
0&=\frac{d^2}{dt^2}\frac{\partial L}{\partial v^{\alpha}}+\mathcal{C}_{\alpha\beta}^{\gamma}y^{\beta}\frac{d}{dt}\left(\frac{\partial L}{\partial v^{\alpha}}\right)-\frac{d}{dt}\frac{\partial L}{\partial y^{\alpha}}-\mathcal{C}_{\alpha\beta}^{\gamma}y^{\beta}\frac{d}{dt}\left(\frac{\partial L}{\partial y^{\alpha}}\right)+\rho_{\alpha}^{i}\frac{\partial L}{\partial x^{i}}.
 \end{align*}\end{proposition}

\subsubsection{Extension to constrained systems}
Let $A$ be a Lie algebroid of rank $n$ over a manifold $M$ of dimension $m$; $E$ the prolongation Lie algebroid of $A$ over $\tau_{A}:A\to M$ with rank $2n$ and $L:A^{(2)}\ra\R$ be a Lagrangian function defined on the set of admissible elements. Consider $\mathcal{M}\subset A^{(2)}$, an embedded submanifold of $A^{(2)}$ given by the vanishing of $\bar{m}$ independent constraints functions $\Psi^{A}:\mathcal{M}\ra\R, \quad A=1,\ldots,\bar{m}.$ Assume the constraints can be written as
$$v^{A}=\Psi^{A}(x^{i},y^{\alpha},v^{a}), 1\leq A\leq\bar{m};\quad \bar{m}+1\leq a\leq n;
1\leq\alpha\leq n.$$ 
Then, $(x^{i},y^{\alpha},v^{a})$ are local adapted coordinates for the submanifold $\mathcal{M}$ of $A^{(2)}.$ Denote by $i_{\mathcal{M}}:\mathcal{M}\ra E$ the inclusion on $E$ given by $$i_{\mathcal{M}}(x^{i},y^{\alpha},v^{a})=(x^{i},y^{\alpha},b^{\alpha},v^{a},\Psi^{A}(x^{i},y^{\alpha},v^{a})).$$


As with unconstrained systems, we can construct the submanifold of $(\mathcal{T}^{\tau_{A}^{1}}E)^{*}$ as
$$\Sigma_{\mathcal{M}}=\{\mu\in(\mathcal{T}^{\tau_{A}^{1}}E)^{*}\mid
i_{\mathcal{M}}^{*}\mu=d^{N}L\}.$$
Locally, it is determined by
\begin{align*}
\mu_{i}^{(1)}&=\rho_{\alpha}^{i}\frac{\partial L}{\partial x^{i}}-\mu_{A}^{(4)}\rho_{\alpha}^{i}\frac{\partial\Psi^{A}}{\partial x^{i}},\\
\mu_{\alpha}^{(2)}+\mu_{\alpha}^{(3)}&=\frac{\partial L}{\partial y^{\alpha}}-\mu_{A}^{(4)}\frac{\partial\Psi^{A}}{\partial y^{\alpha}},\\
\mu_{a}^{(4)}&=\frac{\partial L}{\partial v^{a}}-\mu_{A}^{(4)}\frac{\partial L}{\partial v^{a}}.
\end{align*}
Therefore it is possible to construct the Lagrangian submanifold
$$
S_{L}=\alpha_{E}^{-1}(\Sigma_{\mathcal{M}})=\{x\in\mathcal{T}^{\tau_{A^{*}}^{1}}E\mid\alpha_{E}^{-1}(x)\in\Sigma_{\mathcal{M}}\},
$$
and following the same procedure as before, it is possible to obtain the following system of differential equations 
\begin{align*}
l_{\alpha}+\mathcal{C}_{\alpha\beta}^{\gamma}p_{\gamma}q^{\beta}&=\rho_{\alpha}^{i}\frac{\partial L}{\partial x^{i}}-\mu_{A}^{(4)}\rho_{\alpha}^{i}\frac{\partial\Psi^{A}}{\partial x^{i}},\\
\bar{l}_{\alpha}+p_{\alpha}&=\frac{\partial L}{\partial y^{\alpha}}-\mu_{A}^{(4)}\frac{\partial\Psi^{A}}{\partial y^{\alpha}},\\
\bar{p}_{\alpha}&=\frac{\partial L}{\partial v^{a}}-\mu_{A}^{(4)}\frac{\partial\Psi^{A}}{\partial v^{a}}.
\end{align*}
Differentiating with respect the time the last equation, using the condition $\dot{\bar{p}}_{\alpha}=\bar{l}_{\alpha}$ and replacing in the second equation we obtain that
$$
p_{\alpha}=\frac{\partial L}{\partial y^{\alpha}}-\mu_{A}^{(4)}\frac{\partial\Psi^{A}}{\partial y^{\alpha}}-\frac{d}{dt}\left(\frac{\partial L}{\partial v^{a}}\right)+\dot{\mu}_{A}^{(4)}\frac{\partial\Psi^{A}}{\partial v^{a}}+\mu_{A}^{(4)}\frac{d}{dt}\left(\frac{\partial\Psi^{A}}{\partial v^{a}}\right)
$$
and using that $\dot{p}_{\alpha}=l_{\alpha}$ we have that
\begin{align*}\label{vakonomic}
0=&\frac{d}{dt}\left(\frac{\partial L}{\partial y^{\alpha}}\right)-\dot{\mu}_{A}^{(4)}\frac{\partial\Psi^{A}}{\partial y^{\alpha}}-\mu_{A}^{(4)}\frac{d}{dt}\left(\frac{\partial\Psi^{A}}{\partial y^{\alpha}}\right)-\frac{d^2}{dt^2}\left(\frac{\partial L}{\partial v^{a}}\right)+\ddot{\mu}_{A}^{(4)}\frac{\partial\Psi^{A}}{\partial v^{a}}
\nonumber
\\&-\rho_{\alpha}^{i}\frac{\partial L}{\partial x^{i}}+2\dot{\mu}_{A}^{(4)}\frac{d}{dt}\left(\frac{\partial\Psi^{A}}{\partial v^{a}}\right)+\mu_{A}^{(4)}\frac{d^2}{dt^2}\left(\frac{\partial\Psi^{A}}{\partial v^{a}}\right)+\mathcal{C}_{\alpha\beta}^{\gamma}y^{\beta}\frac{\partial L}{\partial y^{\alpha}}\\&-\mathcal{C}_{\alpha\beta}^{\gamma}y^{\beta}\mu_{A}^{(4)}\frac{\partial\Psi^{A}}{\partial y^{\alpha}}-\mathcal{C}_{\alpha\beta}^{\gamma}y^{\beta}\frac{d}{dt}\left(\frac{\partial L}{\partial v^{a}}\right)+\mathcal{C}_{\alpha\beta}^{\gamma}y^{\beta}\dot{\mu}_{A}^{(4)}\frac{\partial\Psi^{A}}{\partial v^{a}}\\&+\mathcal{C}_{\alpha\beta}^{\gamma}y^{\beta}\mu_{A}^{(4)}\frac{d}{dt}\left(\frac{\partial\Psi^{A}}{\partial v^{a}}\right)+\mu_{A}^{(4)}\rho_{\alpha}^{i}\frac{\partial\Psi^{A}}{\partial x^{i}}.\nonumber
\end{align*}

The former equations together equations $$\displaystyle{\rho_{\alpha}^{i}y^{\alpha}=\frac{dx^{i}}{dt}, v^{a}=\frac{dy^{a}}{dt},  v^{A}=\Psi^{A}(x^{i},y^{\alpha},v^{a})}$$ are the second-order constrained Euler-Lagrange equations for $L:A^{(2)}\to\mathbb{R}$.

\section{Examples}

\begin{example}
Consider the Lie algebroid $TM,$with vector bundle projection denoted by $\tau_{TM}:TM\ra M.$ The prolongation of $TM$ over $\tau_{TM}$ is $E=TTM$. The prolongation of $E$ over the bundle projection $\tau_{(TTM)^*}:(TTM)^{*}\ra TM$ is $T(T^{*}TM)$ and $TE$ is a Lie algebroid over $E$. If $N\subset TTM$ is a submanifold of $TTM,$ then $\tau_{T(TTM)}\mid_{TN}:TN\ra N$ is a Lie subalgebroid over $N.$ The set of admissible elements is identified with the second order tangent bundle $T^{(2)}M.$  Therefore, considering $N=T^{(2)}M$ with local coordinates $(x^i,y^i,v^{i})$,  $T(T^{(2)}M)$ is a Lie subalgebroid over $T^{(2)}M$. 
    
Local coordinates on $T^{*}(TTM)$ denoted by $(x^{i},y^{i},\dot{x}^{i},\dot{y}^{i},p^{x}_i,p^{y}_i,p^{\dot{x}}_i,p^v_i)$. The vector bundles isomorphisms $ \alpha_{TTM}:T(T^{*}TM)\ra T^{*}(TTM)$ and $\beta_{(TTM)^*}:T(T^{*}TM)\ra T^{*}(T^{*}TM)$ are given by 
$$
\alpha_{TTM}\left(x^{i},y^{i},p^{x}_i,p^{y}_i,\dot{x}^i,\dot{y}^{i},\dot{p}^{x}_i,\dot{p}^{y}_i\right)= (x^{i},y^{i},\dot{x}^i,\dot{y}^{i},p^{x}_i,p^{y}_i, \dot{p}^{x}_i,\dot{p}^{y}_i);
$$
$$ 
\beta_{(TTM)^*}\left(x^{i},y^{i},p^{x}_i,p^{y}_i,\dot{x}^i,\dot{y}^{i},\dot{p}^{x}_i,\dot{p}^{y}_i\right)=(x^{i},y^{i},p^{x}_i,p^{y}_i,-\dot{p}^{x}_i,-\dot{p}^{y}, \dot{x}^{i},\dot{y}^{i}).
$$

The Lagrangian Lie subalgebroid $\Sigma_{T^{(2)}M}$ is given by
$$
\Sigma_{T^{(2)}M}=\{\mu\in T^{*}(TTM)\mid i^{*}_{TN}\mu=d^{N}L\}\subset T^{*}(TTM)
$$ 
where $\mu=(x^{i},y^{i},\dot{x}^i,\dot{y}^{i},p^{x}_i,p^{y}_i, \dot{p}^{x}_i,\dot{p}^{y}_i)$ and 
 $i_{TN}:TN\ra T(TTM)$ is the canonical inclusion. Therefore, $$S_{L}=\alpha_{TTM}^{-1}(\Sigma_{T^{(2)}M})=\{x\in T(T^{*}TM)\mid \alpha_{TTM}(x)\in\Sigma_{T^{(2)}M}\}$$ is a Lagrangian submanifold of $T(T^{*}TM)$ and $\Sigma_{T^{(2)}M}$ is locally characterized by the equation
$$ 
p^{x}_i=\frac{\partial L}{\partial x^{i}},\quad p^{y}_i+\dot{p}^{x}_i=\frac{\partial L}{\partial {y}^{i}},\quad \dot{p}^{y}_i=\frac{\partial L}{\partial v^{i}}.$$
Proceeding as in Section \ref{section3.3}, we recover the geometric formulation of the second-order Euler-Lagrange equations obtained by de Le\'on and Lacomba in \cite{LeonLacomba}. In particular, the second-order Euler-Lagrange equations reduce to
$$
0=\frac{\partial L}{\partial x^{i}}+\frac{d^{2}}{dt^{2}}\left(\frac{\partial L}{\partial v^{i}}\right)-\frac{d}{dt}\left(\frac{\partial L}{\partial y^{i}}\right),\quad
y^{i}=\dot{x}^{i},\quad v^{i}=\frac{d}{dt}\dot{x}^{i}.
$$
Assume we have some constraint functions $v^{A}=\Psi^{A}(x^{i},y^{\alpha},v^{a})$ ,at this point, it  is a direct computation to corroborate that the second-order constrained equations are 
\begin{align*}
0=&\frac{d}{dt}\left(\frac{\partial L}{\partial y^{A}}\right)-\dot{\mu}_{\alpha}^{(4)}\frac{\partial\Psi^{\alpha}}{\partial y^{A}}-\mu_{\alpha}^{(4)}\frac{d}{dt}\left(\frac{\partial\Psi^{\alpha}}{\partial y^{A}}\right)-\frac{d^2}{dt^2}\left(\frac{\partial L}{\partial v^{a}}\right)+\ddot{\mu}_{\alpha}^{(4)}\frac{\partial\Psi^{A}}{\partial v^{a}}\\&+2\dot{\mu}^{(4)}_{\alpha}\frac{d}{dt}\left(\frac{\partial\Psi^{\alpha}}{\partial v^{a}}\right)+\mu_{\alpha}^{(4)}\frac{d^2}{dt^2}\left(\frac{\partial\Psi^{\alpha}}{\partial v^{a}}\right)-\frac{\partial L}{\partial x^{A}}+\mu_{\alpha}^{(4)}\frac{\partial\Psi^{A}}{\partial x^{A}},
\end{align*} 
together with $\displaystyle{v^{\alpha}=\Psi^{\alpha}(x^{A},y^{A},v^{a}), y^{i}=\frac{dx^{i}}{dt},v^{a}=\frac{dy^{a}}{dt}}$.
\end{example}

\begin{example}
Consider the Lie algebroid given by a finite dimensional real Lie algebra $\mathfrak{g}$ over a point $g,$ where $\tau_{\mathfrak{g}}:\mathfrak{g}\ra\{g\}$ denotes the bundle projection. The fiber of the prolongation $\mathcal{T}^{\tau_{\mathfrak{g}}}\mathfrak{g}$ which contain the element $\xi\in\mathfrak{g}$ is $\mathfrak{g}\times T_{\xi}\mathfrak{g},$ that is, $\displaystyle{\mathcal{T}^{\tau_{\mathfrak{g}}}\mathfrak{g}=\bigcup_{\xi\in\mathfrak{g}}(\mathfrak{g}\times T_{\xi}\mathfrak{g})}$ and the structure of the Lie algebroid
$(\mathcal{T}^{\tau_{\mathfrak{g}}}\mathfrak{g}, 
\lcf\cdot,\cdot
\rcf, \rho_{\mathfrak{g}})$ over $\mathfrak{g}$ is characterized by the bracket of sections and anchor map: 
$\lcf(\xi,X),(\eta,Y)\rcf=([\xi,\eta]_{\mathfrak{g}},[X,Y]),$ and $\rho_{\mathfrak{g}}(\xi,X)=X$ for any $\xi,\eta\in\mathfrak{g}$ and $X,Y\in\mathfrak{X}(\mathfrak{g}).$

Given a basis of section on $\mathfrak{g}$ it is possible to induce local coordinates $(\xi_1,\xi_2,\xi_3)$ on the prolongation Lie algebroid, where  $\displaystyle{\mathcal{T}^{\tau_{\mathfrak{g}}}\mathfrak{g}=\mathfrak{g}\times\mathfrak{g}\times\mathfrak{g}=:3\mathfrak{g}.}$ The bundle projection is denoted by $\tau_{\al}^{1}$ and defined to be the projection onto the first factor; and the anchor map $\rho^1_{\mathfrak{g}}:3\mathfrak{g}\ra T\mathfrak{g}$ is given by $\rho^{1}_{\mathfrak{g}}(\xi_1,\xi_2,\xi_3)=(\xi_1,\xi_3).$ The set of admissible elements is given by $\mathfrak{g}^{(2)}=\mathfrak{g}\times\mathfrak{g}=:2\mathfrak{g},$ where the inclusion $i_{3\mathfrak{g}}:2\mathfrak{g}\ra 3\mathfrak{g}$ is $i_{3\mathfrak{g}} (\xi,\eta)=(\xi,\xi,\eta)$. 

Note that $E^{*}=\mathfrak{g}^{*}$ and  denoting by $\tau_{\mathfrak{g}^{*}}:\mathfrak{g}^{*}\to\{g\}$ we have $\mathcal{T}^{\tau_{\mathfrak{g}^{*}}}\mathfrak{g}=\mathfrak{g}\times\mathfrak{g}^{*}\times\mathfrak{g}^{*}=:\mathfrak{g}\times 2\mathfrak{g}^{*}$, a Lie algebroid over $\mathfrak{g}^{*}$, with bundle projection determined by the projection onto the second factor, that is, $\tau^1_{\mathfrak{g}^{*}}:\mathfrak{g}\times 2\mathfrak{g}^{*}\ra \mathfrak{g}^{*}$ is given by $\tau^1_{\mathfrak{g}^{*}}(\xi,\mu_1,\mu_2)=\mu_2$. Note also that $\displaystyle{\mathcal{T}^{\tau^{1}_{\mathfrak{g}}}(\mathcal{T}^{\tau_{\mathfrak{g}}}\mathfrak{g})}\simeq 7\mathfrak{g}$ is a Lie algebroid over $3\mathfrak{g}$.  Lemma \ref{th1} tell us that by considering the submanifold of $3\mathfrak{g}$ determined by the set of admissible elements $2\mathfrak{g}$, then $A_N=\rho_2^{-1}(T(2\mathfrak{g}))$ is a Lie subalgebroid over $2\mathfrak{g},$ where $\rho_2:7\mathfrak{g}\ra 3\mathfrak{g}\times T(3\mathfrak{g})$ denotes the anchor map of the Lie algebroid $7\mathfrak{g}$ over $3\mathfrak{g}.$
Consider the following identifications that locally provide \eqref{eqq1} and \eqref{eqq2} $\left(\mathcal{T}^{\tau^{1}_{\mathfrak{g}}}(3\mathfrak{g})\right)^{*}\simeq 3\mathfrak{g}\times 4\mathfrak{g}^{*}$, 
$\mathcal{T}^{\tau^{1}_{\mathfrak{g}^{*}}}(3\mathfrak{g})\simeq(\mathfrak{g}\times 2\mathfrak{g}^{*})\times(2\mathfrak{g}\times 2\mathfrak{g}^{*})$, and $\left(\mathcal{T}^{\tau^{1}_{\mathfrak{g}^{*}}}(3\mathfrak{g})\right)^{*}\simeq(2\mathfrak{g}^{*}\times \mathfrak{g})\times(2\mathfrak{g}^{*}\times 2\mathfrak{g})$.

Denoting by $\bar{r}=(\xi,p,\bar{p},q,\bar{q},l,\bar{l})$ an element of $(\mathfrak{g}\times 2\mathfrak{g}^{*})$ the Tulczyjew's symplectomorphism is given by $\alpha_{3\mathfrak{g}}(\bar{r})=(\xi,l+\hbox{ad}^{*}_{q}p,\bar{l},q,\bar{q},p,\bar{p})\in (2\mathfrak{g}^{*}\times\mathfrak{g})\times(2\mathfrak{g}^{*}\times 2\mathfrak{g}).$ Therefore, for $\bar{\gamma}=((\mu_1,\mu_2,\xi_1),(\mu_3,\mu_4,\xi_2,\xi_3))\in(2\mathfrak{g}^{*}\times\mathfrak{g})\times(2\mathfrak{g}^{*}\times 2\mathfrak{g})$, $$\alpha_{3\mathfrak{g}}^{-1}(\bar{\gamma})=(\xi_1,\mu_3-\hbox{ad}^{*}_{\xi_3}\mu_1,\mu_4,\mu_2,\xi_2,\xi_3,\mu_1,\mu_2).$$ 
 
 The dynamics on $\Sigma_{2\mathfrak{g}}\subset (2\mathfrak{g}^{*}\times\mathfrak{g})\times(2\mathfrak{g}^{*}\times 2\mathfrak{g})$ is locally described by 

$$\mu_3=\hbox{ad}_{\xi_3}^{*}\mu_1,\quad\mu_4=\frac{\partial L}{\partial\xi_1}-\mu_1,\quad\mu_2=\frac{\partial L}{\partial\xi_2}.$$
 
Proceeding as in Section \ref{section3.3}, and using the identification $\xi_1=\xi_3$ the second-order Euler-Lagrange equations for a Lagrangian $L:2\mathfrak{g}\to\mathbb{R}$ are $$\frac{d^2}{dt^2}\left(\frac{\partial L}{\partial\xi_2}\right)-\frac{d}{dt}\left(\frac{\partial L}{\partial\xi_1}\right)=\hbox{ad}_{\xi_1}^{*}\left(\frac{d}{dt}\left(\frac{\partial L}{\partial\xi_2}\right)\right)-\hbox{ad}_{\xi_1}^{*}\left(\frac{\partial L}{\partial\xi_1}\right).$$
\end{example}

\section*{Appendices} 

\subsection*{Appendix A:}  The aim of this Appendix consists on discuss the existence of a Lagrangian Lie subalgebroid used in Section $3.3$. More precisely,  if $(A,\lcf\cdot,\cdot\rcf,\rho)$ is a Lie algebroid over $M$, the submanifold $\Sigma_{A_N}\subset (\mathcal{T}^{\tau_A^{1}}E)^{*}$, given by $$
\Sigma_{A_N}=\{\mu\in(\mathcal{T}^{\tau_A^{1}}E)^{*} \mid i^{*}_{A_N}\mu=d^{A_N}L\},
$$
where $A_N=\rho_{\tau_{A}^{2}}^{-1}(TN),$ $i_{A_N}:A_N\rightarrow\mathcal{T}^{\tau_A^{1}}E$ is the canonical inclusion and $N=A^{(2)}$ is a  Lagrangian Lie subalgebroid of $(\mathcal{T}^{\tau_A^{1}}E)^{*}$. 

First we recall the construction done in Section $7$ \cite{LMM} and then we apply it to our situation.

Denote $(x^i)$ local coordinates on $M$, $i=1,\ldots,n=dim(M)$ $\{e_{\alpha}\}$ with $\alpha=1,\ldots,m=\hbox{rank}(A)$ a basis of sections of $A$ and $\{e^{\alpha}\}$ its dual, a basis of $A^{*}$ over $M$. In this basis, the differential of a smooth function on $M$, $f:M\to\mathbb{R}$ is given by $d^{A}f=\frac{\partial f}{\partial x^{i}}\rho_{\alpha}^{i}e^{\alpha}\in\Gamma(A^{*})$. Both basis induce local coordinates on $A$ and $A^{*}$ respectively, denoted by $(x^i,y^{\alpha})\in A$ and $(x^{i},p_{\alpha})\in A^{*}$ respectively.

Define the submanifold of $A^{*}$ determined by $S=\hbox{Im}(d^{A}f)$. Local coordinates on $S$ are induced by the basis $\{e^{\alpha}\}$ as $\displaystyle{\left(x^{i},\frac{\partial f}{\partial x^{i}}\rho_{\alpha}^{i}(x)\right)\in A^{*}}$ describing locally the submanifold $S$ as $$S=\hbox{Im}(d^{A}f)=\Big{\{}(x^i,p_{\alpha})\in A^{*}|p_{\alpha}=\frac{\partial f}{\partial x^{i}}\rho_{\alpha}^{i}\Big{\}}$$

 From Section \ref{s2} we know that The prolongation $\mathcal {T}^{\tau_{{A}^{*}}}A$ of $A$ over the vector bundle projection $\tau_{{A}^{*}}:A^*\to M$ is a symplectic Lie algebroid. Besides this, if $x\in M$ then $j_x:TA_x^*\to \mathcal {T}^{\tau_{{A}^{*}}}A$, $i_x:A^*_x\to A^*$ is a Lagrangian Lie subalgebroid of $\mathcal {T}^{\tau_{{A}^{*}}}A$, where $j_x$ and $ i_x$  are defined  by
$$ j_x(v)=(0(x),v),\quad i_x(\alpha)=\alpha,$$
for $v\in T A^*_x$, $\alpha\in A_x^*$, with $0:M\to A$  the zero section of $A$ (see \cite{LMM}, Section $7$). 

For $\gamma\in \Gamma(A^*)$, denote by $F_\gamma$ the vector bundle over $\gamma(M)$ given by
\begin{equation}\label{fgamma}
F_\gamma=\{(b,(T\gamma)(\rho(b)))\in A\times TA^*\mid b\in A\}
\end{equation}
and by $j_\gamma:F_\gamma\to \mathcal {T}^{\tau_{{A}^{*}}}A$ and $i_\gamma:\gamma(M)\to A^*$ the canonical inclusions. The pair $((Id,T\gamma\circ \rho),\gamma)$ is an isomorphism between the vector bundles $A$ and $F_\gamma$, where the map $(Id,T\gamma\circ \rho):A\to F_\gamma$ is given by
$$(Id,T\gamma\circ \rho)(b)=(b,T\gamma(\rho(b))),\quad \mbox{ for }b\in A.$$
Hence, $F_\gamma$ is a Lie algebroid over $\gamma(M)$. In \cite{LMM} (Proposition 7.3) it was shown that if $\gamma\in \Gamma(A^*)$ then $j_\gamma:F_\gamma\to\mathcal {T}^{\tau_{{A}^{*}}}A$, $i_{\gamma}:\gamma(M)\to A^*$ is a Lagrangian Lie subalgebroid over $\gamma(M)\subset A^{*}$ of the symplectic Lie algebroid $\mathcal {T}^{\tau_{{A}^{*}}}A$ over $A^{*}$ if and only if $d^{A}\gamma=0.$

Next we will apply the previous construction and the result in Proposition 7.3 of \cite{LMM} to our situation.  Consider the map $\gamma=d^{A}f\in\Gamma(A^{*})$, then $\gamma(M)=\hbox{Im}d^{A}f=\hbox{Im}\gamma$. The map $T\gamma:TM\to TS=T( \hbox{Im}d^{A}f)$ is given by $$T\gamma(x^{i},\dot{x}^{i})=\left(x^i,\rho_{\alpha}^{i}\frac{\partial f}{\partial x^{i}},\dot{x}^{i},\frac{\partial^2f}{\partial x^i\partial x^{j}}\dot{x}^{j}\rho_{\alpha}^{i}+\frac{\partial f}{\partial x^{i}}\frac{\partial\rho_{\alpha}^{i}}{\partial x^{j}}\rho_{\beta}^{j}y^{\beta}\right).$$ Therefore $F_{\gamma}$ is locally described by \begin{align*}F_{\gamma}=&\Big{\{}(x^i,y^{\alpha},p_{\alpha},\bar{p}_{\alpha})\in \mathcal{T}^{\tau_{A^{*}}}A|\\ &\qquad\qquad\qquad p_{\alpha}=\rho_{\alpha}^{i}\frac{\partial f}{\partial x^{i}},\bar{p}_{\alpha}=\frac{\partial^2f}{\partial x^i\partial x^{j}}\dot{x}^{j}\rho_{\alpha}^{i}+\frac{\partial f}{\partial x^{i}}\frac{\partial\rho_{\alpha}^{i}}{\partial x^{j}}\rho_{\beta}^{j}y^{\beta}\Big{\}}.\end{align*}

Since $d^{A}\gamma=(d^{A})^2f=0$ by Proposition 7.3, \cite{LMM} $F_{(\hbox{Im}d^{A}f)}$ is a Lagrangian Lie subalgebroid over $S$ of the symplectic Lie algebroid $\mathcal{T}^{\tau_A^{*}}A$. This also implies that $\hbox{rank} (F_{\gamma})=\frac{1}{2}\hbox{rank}(\mathcal{T}^{\tau_{A^{*}}}A)$ and $\Omega_{i(a^{*})}\mid_{j(F_{\gamma})\times j(F_{\gamma)}}=0$ for $a^{*}\in S=\gamma(M)=\hbox{Im}d^{A}f$.

By Lemma \ref{th1} if $N$ is a submanifold of $M$, $A_N=\rho^{-1}(TN)$ is a Lie algebroid over $N$. Consider local coordinates $(x^{j})$ on $N$ with $j=1,\ldots,k<n$ and $i_{N}(x^j)=(x^j,x^{I}=0)\in M$, with $I=1,\ldots,n-k$. 

Denote by $\{\sigma_a\}$ a basis of section for $A_N$ and extend this to $M$ such that $e_a:M\to A$ satisfies $e_{a}|_{N}=\sigma_a$. Therefore, we can complete this basis to a basis of section for $A$, $\{e_a,e_{A}\}=\{e_\alpha\}$. This basis of sections induce local coordinates $(x^j,x^{I},y^{a},y^{A})$ on $A$. Therefore, $A_N$ is locally described by $(x^j,x^{I}=0,y^{a},y^{A}=0)$, that is, $(x^j,y^a)$ are local coordinates on $A_N$. 

For $f\in\mathcal{C}^{\infty}(M)$ we have that $d^{A_N}f=\frac{\partial f}{\partial x^j}\rho_a^{j}e^a$ where $\{e^a,e^{A}\}$ is the dual basis of $\{e_a,e_A\}$. The basis $\{e^a,e^A\}$ indeces local coordinates $(x^j,p_a)$ on $A^{*}_{N}$. Then, if we denote by $\widetilde{S}=\hbox{Im } d^{A_N}f$, $\widetilde{S}$ is locally described by $$\widetilde{S}=\Big{\{}(x^j,p_a)\in A_N^{*}|p_a=\frac{\partial f}{\partial x^j}\rho_a^j\Big{\}}\in A_{N}^{*},$$ with canonical inclusion on $A^{*}$ determined by $(x^j,0,\rho_a^j\frac{\partial f}{\partial x^j},p_{A})\in A^{*}$. 

Consider the prolongation of $A$ by $\tau_{A_{N}^{*}}:A_N^{*}\to N$ $$\mathcal{T}^{\tau_{A_{N}^{*}}}A=\Big{\{}(b,p_{a^{*}})\in A\times T_{a^{*}}A_{N}^{*}|\rho(b)=(T\tau_{A_N^{*}})(p_{a^*})\Big{\}},$$ a Lie algebroid over $A_N^{*}$.  Given an element $a^{*}\in A^{*}$, $a^{*}=e_a^{*}e^{a}+e_{A}^{*}e^{A}$ we construct the basis of section $\{\tilde{e}_a,\tilde{e}_{A},\bar{e}_a,\bar{e}_A\}$ by 
\begin{align*}
\tilde{e}_a(a^*)&=\left(e_a^{*},e_{\alpha}(\tau_{A^{*}}(e_a^*)),\rho_{\alpha}^{j}\frac{\partial}{\partial x^j}\Big{|}_{a^{*}}\right),\quad \tilde{e}_A(a^*)=(e_{A}^{*},e_{\alpha}(\tau_{A^{*}}(e_{A}^{*})),0)\\
\bar{e}_{a}(a^{*})&=\left(e_a^{*},0,\frac{\partial}{\partial p_a}\Big{|}_{a^{*}}\right),\quad \bar{e}_{A}(a^{*})=\left(e_A^{*},0,\frac{\partial}{\partial p_A}\Big{|}_{a^{*}}\right).\end{align*}

Denoting by $\{\tilde{e}^{a},\tilde{e}^{A},\bar{e}^{a},\bar{e}^{A}\}$ the dual basis of the basis given before, a basis of $(\mathcal{T}^{\tau_{A_N}^{*}}A)^{*}$ we have that the Liouville section is determined by $\lambda_{A_N}(x^j,x^I,p_a,p_A)=p_a\tilde{e}^{a}+p_A\bar{e}^{A}$ and the canonical symplectic section of 
the Lie algebroid $\mathcal{T}^{\tau_{A_N}^{*}}A$ is locally described by $$\Omega_{A_N}=-d^{R}\lambda_{A_N}=\tilde{e}^{a}\wedge\bar{e}^{a}+\tilde{e}^{a}\wedge\bar{e}^{a}+\tilde{e}^{a}\wedge\bar{e}^{A}+\frac{1}{2}\mathcal{C}_{ab}^{c}p_c\tilde{e}^{a}\wedge\tilde{e}^{b}$$
where $R=\mathcal{T}^{\tau_{A_N}^{*}}A$.
Note that  $d^{R}f=\rho_a^{j}\frac{\partial f}{\partial x^j}\tilde{e}^{a}+\frac{\partial f}{\partial p_a}\bar{e}^{a}+\frac{\partial f}{\partial p_A}\bar{e}^{A}$, $f\in\mathcal{C}^{\infty}(A_N^{*})$, $d^{R}\tilde{e}^{a}=-\frac{1}{2}\mathcal{C}_{ab}^{c}\tilde{e}^{a}\wedge\tilde{e}^{b}$, $d^{R}\tilde{e}^{A}=-\frac{1}{2}\mathcal{C}_{AB}^{c}\tilde{e}^{A}\wedge\tilde{e}^{B}$, $d^{R}\bar{e}^{a}=0$, $d^{R}\bar{e}^{A}=0$ and $d^{R}\Omega_{A_N}=0$ because $d^2=0$ since $R$ is a Lie algebroid over $A_{N}^{*}$. Therefore  $\mathcal{T}^{\tau_{A_N}^{*}}A$ is a symplectic Lie algebroid. Next, defining $$F_S=\Big{\{}(b,w)\in A\times TA^{*}|w\in TS, b\in A_N, \rho(b)=T\tau_{A_N^{*}}(w)\Big{\}}\subset\mathcal{T}^{\tau_{A_N^{*}}}A$$ where $S=\{a\in A^{*}|i_{A_N}^{*}(a)\in\hbox{ Im }d^{A_N}f\}\subset A^{*}$ is a Lagrangian Lie subalgebroid by Proposition 7.3, \cite{LMM}. By choosing $N=A^{(2)}$, $f=L:A^{(2)}\to\mathbb{R}$ and $A$ before as $\mathcal{T}^{\tau^{1}_{A}}(\mathcal{T}^{\tau_A}A)$, $\Sigma_{A_N}$ is a Lagrangian Lie subalgebroid.

\subsection*{Appendix B: Implicit differential equation on Lie algebroids}
\label{appendixb}

An \textit{implicit differential equation} on a general Lie algebroid $(A,\lcf\cdot,\cdot\rcf,\rho)$ over $M$ is a submanifold $S\subset A$ (not necessarily a vector subbundle). A \textit{solution} of $S$ is any curve $\gamma: I\subset\R\ra A$, such that $(\gamma(t),\dot\gamma(t))\in S\times T_{\gamma(t)}S$, for all $t\in I$ satisfying $\frac{d}{dt}(\tau_{A}\circ\gamma)=\rho\circ\gamma$, where $\tau_A:A\rightarrow M$ is the vector bundle projection. The implicit differential equation $S$ will be said to be {\it integrable at a point} if there exists a solution $\gamma$ of $S$ such that the tangent curve passes through it. Furthermore, the implicit differential equation will be said to be {\it integrable} if it is integrable at {\it all points}. The integrable part of an  implicit differential equation  $S$ is the subset of all integrable points of $S$. The {\it integrability problem} consists in identifying such subset.

A recursive algorithm  that allows to extract the integrable part of an implicit differential equation $S$ (possible empty) was presented in \cite{singular} and extended in the context of Lie algebroids in \cite{IMMS}. We shall define the subsets $S_0=S,\,\,\,\,C_0=C$, and recursively for every $k\geq1$,
$$ S_k=S_{k-1}\cap \rho^{-1}(TC_{k-1}),\quad C_k=\tau_A(S_k), $$
then, eventually the recursive construction will \textit{stabilize} in the sense that $S_k=S_{k+1}=...=S_{\infty}$, and $C_k=C_{k+1}=...=C_{\infty}$ (Note that by construction that $S_{\infty}\subset TC_{\infty}$). Therefore, provided that the adequate regularity conditions are satisfied during the application of the algorithm, the implicit differential equation $S_{\infty}$ will be integrable and it will solve the integrability problem.



\medskip
\medskip

\end{document}